# Learning with a Network of Competing Synapses

**Ajaz Ahmad Bhat[1], Gaurang Mahajan[1], Anita Mehta[1,2]***

**1** S N Bose National Centre for Basic Sciences, Salt Lake, Calcutta, India, **2** Institut de Physique Théorique, CEA Saclay, Gif-sur-Yvette, France

## Abstract

Competition between synapses arises in some forms of correlation-based plasticity. Here we propose a game theory-inspired model of synaptic interactions whose dynamics is driven by competition between synapses in their weak and strong states, which are characterized by different timescales. The learning of inputs and memory are meaningfully definable in an effective description of networked synaptic populations. We study, numerically and analytically, the dynamic responses of the effective system to various signal types, particularly with reference to an existing empirical motor adaptation model. The dependence of the system-level behavior on the synaptic parameters, and the signal strength, is brought out in a clear manner, thus illuminating issues such as those of optimal performance, and the functional role of multiple timescales.

**Funding:** This work was funded by the Department of Science and Technology, Government of India, under the project "Cognitive Science Initiative." The funders had no role in study design, data collection and analysis, decision to publish, or preparation of the manuscript.

**Competing Interests:** The authors have declared that no competing interests exist.

* E-mail: anita@bose.res.in

## Introduction

Natural neural systems possess the capacity for generating purposeful, relevant and directed behavior in a complex, uncertain and ever-changing environment. At the heart of this capacity is their ability to show adaptive behavioral changes in the face of varying external conditions, to learn efficiently and to retain information reliably in memory. Given that the external sensory world is a complex one and contains a spectrum of dynamic processes spanning a gamut of timescales [1–3], it may be expected on general grounds that the dynamical system underlying information processing in brains must have a multiplicity of spatial and temporal scales built in, to take cognizance of, respond to, and deal with its complex multi-scale environment. On the one hand, that neural mechanisms of adaptation and cognition involve short as well as long timescale dynamic phenomena, is amply evidenced by experimental work [4–7]. On the other, several theoretical studies [8–13] carried out in recent times have explored how the introduction of multiple timescales into computational models of learning and adaptation can affect their functional properties, expanding their capabilities and strengthening links with experiment. To give a few representative examples: the introduction of multiple degrees of plasticity in a model for synaptic "metaplasticity" was shown to achieve a desirable balance between receptivity to new stimuli while remaining immune to degradation of older memories, and to give rise to a power law-type forgetting, not just at the system level [9] (consistent with psychophysical findings), but also at the level of a single synapse [14]; a neural network model [11] for motor learning with two membrane time constants was able to demonstrate efficient motor learning, generating a functional hierarchy of motor movement, with more elaborate actions being composed out of sequences of shorter, elementary building-block motor 'primitives' strung together; an empirical model of motor adaptation [10] that contained mutually coupled slow and fast timescale sub-systems, produced better agreement (in relation to alternative models) with a series of experimental findings on hand-reaching behavior, which included such phenomena as savings, anterograde interference, and adaptation rebound.

While learning is integral to neural systems and functionally beneficial at the level of a single individual, many studies have focused on the collective effects of [simple forms of] individual learning and decision-making, i.e. in populations of interacting individuals, or agents. Such distributed systems, exemplifying social or ecological group behavior [15], also share similarities with interacting systems of statistical physics, in the nature of the local "rules" followed by the individual units as well as in the emergent behavior at the macro level, which can under some circumstances display a high degree of order and coordination [16]. Game-theoretic approaches [17–19] are sometimes brought to bear on such issues, their underlying idea being that the behavior of an individual (its "strategy") is to a large extent determined by what the other individuals are doing. The strategic choices of an individual are thus guided by those of the others, through considerations of the relative "payoffs" (returns) obtainable in interactive games. In this context, a stochastic model of strategic decision-making was introduced in [20], which captures the essence of the above-stated notion, i.e. selection from among a set of *competing* strategies based on a comparison of the *expected* payoffs from them. Depending upon which of the available strategic alternatives (that are being wielded by the other agents) is found to have the most favorable "outcome" in the local vicinity, every individual appropriately revises its strategic choice.

Competition between prevalent strategies and adaptive changes at the individual level characterize the sociologically motivated model of [20]. Given that these two features of competition and adaptation also generally occur across the framework of activity-induced synaptic plasticity, which is the primary mechanism for learning in biological neural systems [21,22], it might be interesting to consider a translation of the notions in [20] to the latter context, as a cross-pollinatory attempt of sorts. In other words, a model for synaptic plasticity that incorporates the brand of competition present in the agent-based strategic learning model could be envisaged. A model was delineated in ref. [23] along these lines, with the types or weights of a plastic synapse taking the place of strategies. This model inherently possesses more than a single timescale, which are interpreted here in terms of the [different] effects of each synapse type on the activation rate of a connected neuron. It turns out to be possible to define a rather simple framework for learning and memory, which involves subjecting the network of interconnected neural units to external signals, and following the changes in the average behavior (in terms of the relative abundances of the different synaptic states) of the system as it responds to the external input. A salient result emerging from the analytical approximation carried out in ref. [23] points to the benefit of choosing disparate synaptic timescales for obtaining longer retention times at the network level. This finding echoes earlier results on two timescales of some theoretical analyses mentioned in the opening paragraph above [10,11]. Of possible and perhaps broader significance is also the fact that in our setup, there is a clear-cut connection between the micro-level physical observables and the macroscopic learning/forgetting rates, at least in the analytically tractable *effective* representation within which we work in [23]. This should allow for a more transparent approach to dealing with questions of optimization of the system performance under different learning protocols, so that behavior is given a microscopically understandable basis.

Taking a cue from the numerical investigations of the motor adaptation model in [10] to which we have alluded in [23], here we shall explore the dynamical outcomes of our model of competitive synaptic interactions under a variety of applied time-dependent external signals. This is expected to reveal in more detail the dependence of the system-level adaptational properties on the synaptic time constants, and thus the model's functional scope as well. Also, given the inherent nonlinearity of this model arising from the synaptic interactions at the basic level, it is reasonable to suppose that it would be better suited to modeling memory. Seen in the context of the linear two-timescale system of ref. [10], which is essentially empirical in nature, our analysis should provide some clues as to the sort of microscopic approach that would be needed towards obtaining an adequate theoretical underpinning of the ideas presented there.

This paper is organized along the following lines: the next subsection provides, by way of background, an outline of the original strategic learning model, followed by an account of the model for synaptic plasticity that adopts elements of the former, and which has been treated in detail in ref. [23]. In the subsequent section, the mean-field representation of the network model is briefly summarized. Working within the limits of this effective description, we illustrate with an example how the choice of synaptic parameters has a bearing on the collective timescales associated with learning and forgetting when the system is subjected to a signal. This idea is elaborated further in Section 4, which extends the foregoing analysis to an analytical-cum-numerical exploration of different forms of input signals and the dynamical responses they elicit. These inputs are meant to reflect, at least in essence, some of the experimental protocols considered in [10]. Finally, the overall picture emerging from our findings is put in perspective, particularly in relation to ref. [10], in Section 5, and potential directions for taking the work further are briefly noted.

### Background: A model for strategic learning

The starting point for our proposal in the present work is the model of competitive learning that was introduced, and analyzed from a physics perspective, in [20]. Although originally intended to describe sociological phenomena like the diffusion of innovations in connected societies, it can be applied just as well to represent any process which involves, at the elemental level, selection from among a set of competing alternatives.

In its original formulation, a distributed population of interacting agents is arranged on the sites of a regular lattice, each being ascribed one of two categories: fast (F) or slow (S). In an evolutionary scenario, for example, these types could stand for two contrasting behavioral strategies prevalent in the population. The type associated with every site is not a given, but can keep changing over time as a function of its nearest neighbors' types; this is where competitive selection plays a role. Thus, every agent/site regularly revises its strategic choice, being guided by a pair of rules: a *majority*-type rule reflecting an inherent tendency to side with the local neighborhood, which is followed by an adaptive *performance-based* rule, involving the selection of the type that is perceived to be locally more successful. The notion of success is measured in terms of the random outcomes of the agents in some "game", with a favorable outcome being ascribed to every F-type (S-type) individual with an *independent* probability $p_+$ (resp. $p_-$). Thus, if an agent is surrounded by $N_+$ ($N_-$) nearest neighbors of the F (S) type, $I_+$ (resp. $I_-$) of which turn out to be successful in a particular trial, the agent arrives at a decision on whether or not to convert by comparing the ratios $I_+/N_+$ and $I_-/N_-$; if, for example, a site is currently associated with the F state, then it will switch to the S type provided that $I_+/N_+ < I_-/N_-$, and remain unchanged otherwise. This rule can be immediately understood by noticing that the ratios just mentioned are nothing but the average payoff per individual ascribed by a site to each of the two types in its neighborhood, assuming of course that success yields a payoff of unity and failure, zero.

It goes without saying that the above outcome-based updates naturally introduce an element of stochasticity into the population dynamics, owing to the random fluctuations inherent in the estimation of the relative payoffs. In ref. [20], a detailed analysis of this model was carried out under the assumption of *coexistence*, i.e. when $p_+ = p_- \equiv p$. Its collective behavior, as a function of the parameter $p$, was shown to exhibit multiple dynamic phases separated by critical phase transitions.

In ref. [23], the notion of competition embodied by the above model has been reformulated in the synaptic context. After all, adaptivity is a feature common to both settings, and it is therefore not unnatural to consider the embedding of the rules of the agent-based learning model into a model for synaptic plasticity, and to understand the implications of doing so for suitably defined learning and memory. This was initiated in ref. [23], and in the present work will be explored in greater detail.

## Results

### Model and effective description

To begin with we shall sketch our model of competitive synaptic interactions that is based on game theory-inspired ideas. We consider a network consisting of neural units connected by undirected, binary, plastic synapses. (It is pertinent to mention here that in working with symmetric, i.e. undirected synapses, we are following in the footsteps of several previous theoretical studies on neural networks (e.g. [24–26]); on the other hand, the binary property approximates synapses with discrete weight states, which also appear in previous modeling work [27,28] and find some experimental support as well [29,30]). Synapses sharing a connected neural unit are treated as mutual neighbors. In a one-dimensional formulation, like the one depicted in Fig. 1, each synapse will thus be associated with two synaptic neighbors. For simplicity the neurons can be represented by binary threshold units, and the two states of the binary synapse, which are interconvertible by definition, are assumed to have different weights, which we label as 'strong' and 'weak' types.

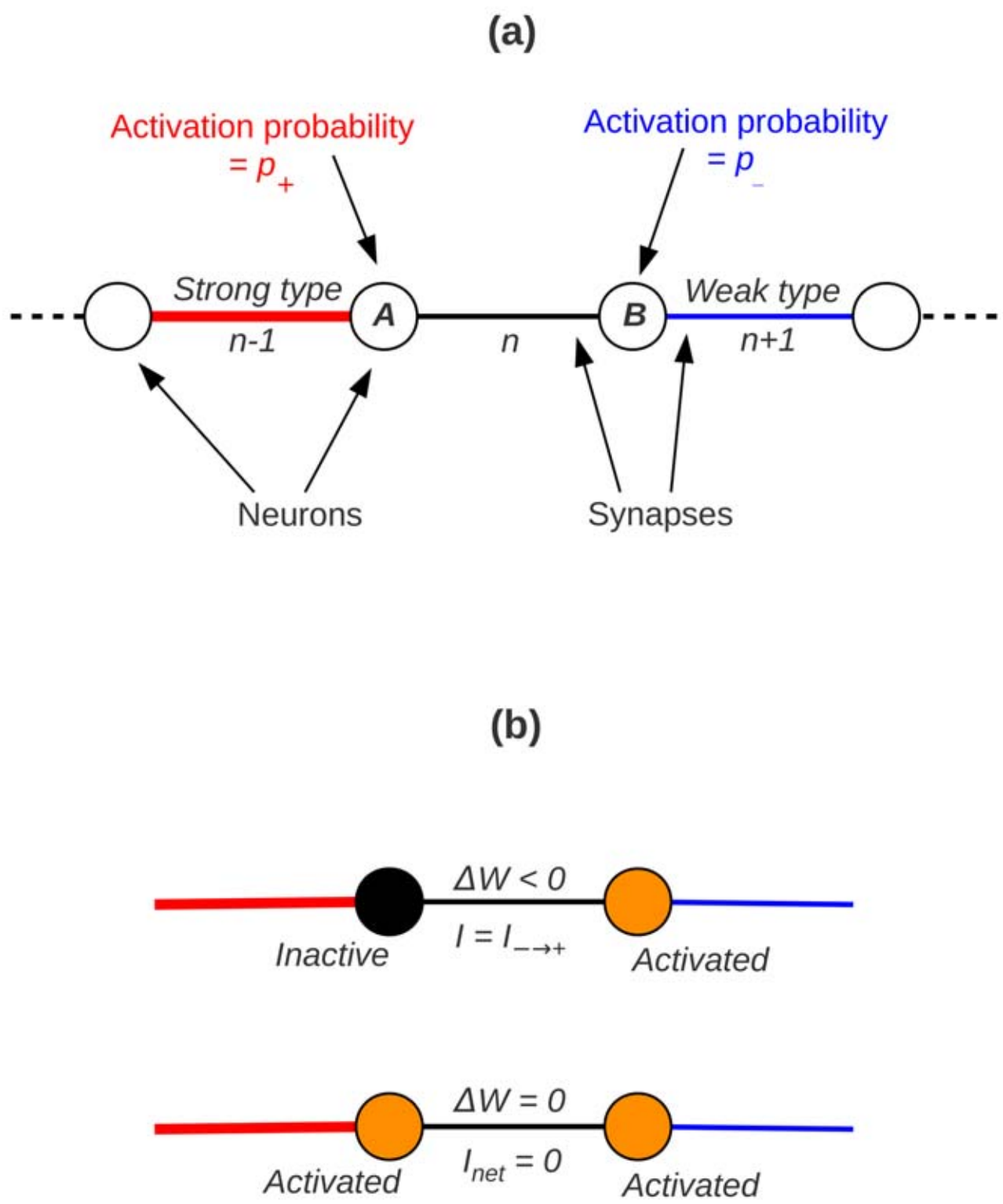

**Figure 1. A model for plastic synapses.** (a) A networked population of binary synapses connecting neurons in a one-dimensional chain. The synapse $n$, under consideration for an update, has a 'strong' type and a 'weak' type neighbor. (b) Two examples of synaptic weight changes (the synaptic configuration is same as above): when neuron B is activated and neuron A is not (upper example), the synaptic current has negative polarity ($I = I_{-\to+}$) and the weight of synapse $n$ is depressed. When both neurons get activated (lower example), the current has net zero polarity ($I_{net} = 0$), and therefore the synaptic weight remains unaffected.

Under the influence of some ongoing neural activity in the network, the synapses undergo plastic switching from one state to the other. In order to motivate the specific plasticity rules that we introduce, we point out that in the configuration shown in Fig. 1, where the middle synapse $n$ is under consideration for a state update, the neurons A and B share this middle synapse in common; thus, in comparing how often the two neurons are found activated, one can factor out the influence of the common synapse, when considering averages, and effectively treat the time-averaged activation frequency of either neuron as being determined only by the single, *other* synapse that the neuron is connected to. This "ignoring of the common denominator" essentially implies that the state of neuron A, say, in Fig. 1 can be considered quite reasonably as an "outcome" to be associated with synapse $n-1$, and similarly with neuron B and synapse $n+1$; thus, neurons can be thought of as taking on the identities of the respective synapses. Recalling the obvious similarity of this situation with that of the abstract model of [20], and taking this analogy further, we set forth the following rules governing the synaptic weight changes, which have an anti-Hebbian flavor: for clarity, we shall first associate with every possible pair of neural outcomes a "polarized" signal, that depends on the states of the two neurons as well as the types of both the adjacent synapses (i.e. the neighbors of the synapse that is being considered for updating at that moment). A positively (negatively) polarized signal is realized when the synapse connects a strong neuron with a weak neuron, *and* the strong (weak) neuron alone is activated. All other possible combinations of neuron types and activation states are associated with zero or unpolarized current. With this definition of polarity, it is proposed that $\Delta w > 0$ ($\Delta w < 0$) whenever there is a positively (negatively) polarized current, and $\Delta w = 0$ in all other cases. Furthermore, to be consistent with the binary nature of the synapses, it is assumed that a strengthening event would effect a weak to strong conversion, while leaving an already strong synapse unchanged (the corresponding logic would hold for a weak synapse). Thus, loosely speaking, the two synapses adjacent to any given synapse "compete" to decide its type, and this continues to happen repeatedly across the entire network. Competition, albeit in other forms, is found to occur in some other models of correlation-based plasticity also [22,31,32].

In order to make some mathematical headway in analyzing such a network of interacting synapses, we shall consider the update dynamics of a single *effective* synapse, that in some sense represents the average state of the whole network. To begin with, in such a picture, the neural outcomes are assumed to be uncorrelated at different locations, and treated as independent random variables, with the probability for activation being obtainable from the time-averaged activation frequency of the neuron. Consistent with the situation described in the previous paragraph, that the effect of the common synapse can be left out on average in comparing the outcomes of its connected neurons, we associate, with each neuron, a probability for activation at any instant that is *only* a function of the other neighboring synapse, being equal to $p_+$ ($p_-$) for a strong (weak) type synapse. Thus, in Fig. 1(b), the probability for activation of neuron A is equal to $p_+$, and that for neuron B is $p_-$; both are independent of the state of the middle synapse that is under consideration for updating.

Having defined these quantities it is rather straightforward to work out the probabilities for potentiating or depressing events to occur at a candidate synapse, given the identities of its neighbors. This is illustrated with an example. Say a synapse has one neighbor of each type, as is depicted in fig. 1. For this configuration, a total of four outcomes for the neuronal pair A–B are possible. There will be no weight changes if both neurons get activated (giving an unpolarized synaptic current) or if both remain silent; the likelihood of this happening is $p_+p_- + (1-p_+)(1-p_-)$. A depressing event ($\Delta w < 0$) occurs if neuron B fires but neuron A remains inactive, and this has a probability $p_-(1-p_+)$. The only remaining possibility is that neuron A gets activated and neuron B does not. This occurs with a probability $p_+(1-p_-)$, and is accompanied by potentiation ($\Delta w > 0$).

Our rules for activity-induced weight changes may have been motivated by game-theoretic notions of strategic competition, but when seen in the light of earlier work on rate-based models of synaptic plasticity, a case can be made for their reasonableness at least in relation to other earlier proposals in the field. In continuous-time models, the firing rate of the neuron, rather than its membrane potential, is taken as the basic dynamical variable, and synaptic plasticity is a continuous process that depends on the firing rates of the pre- and post-synaptic neurons. The dynamical equation describing the time evolution of the synaptic weight usually involves some non-linear function of pre/post-synaptic activities and the weight itself, and in some cases, a dependence on averages of the firing rates over some temporal window has also been motivated [22,33,34]. Drawing on such approaches, we speculate that the rules for synaptic plasticity proposed in the previous paragraph might also be realizable through an iterative, discrete equation symbolically expressed as $\Delta w \propto \phi((u-<u>)(v-<v>))$ [22], where the form of the non-linear function $\phi$ is chosen to provide an appropriate fit to the plasticity rules. Here, $u$ and $v$ represent the activity states of the two connected neurons, which are binary variables in the present set-up, being either active (1) or inactive (0). The symmetric form of the argument of $\phi$ is in keeping with the bi-directional nature of the synapses and ensures that the synaptic response is insensitive to the spatial direction of any current, while still being sensitive to its polarity. The inclusion of time averages of the activity of the connected neurons allows for a characterization of the strengths of the neighboring synapses in this picture, and hence allows for the determination of the polarity of any current at the synapse.

Returning to our previously mentioned intention of having an analytically pliable representation of the system dynamics – even though it could be numerically simulated by using a range of updates, as in the case of the game-theoretic model [20] – we consider a mean-field version of the model. The idea behind the mean-field approximation is that we look at the average behavior in an infinite system. This, at one stroke, deals with two problems: first, there are no fluctuations associated with system size, and second, the approximation that we have made in ignoring the "self-coupling" of the synapse is better realized.

In the mean-field representation, every synapse is assigned a probability (uniform over the lattice) to be either strong ($f_+$) or weak ($f_-$), so that spatial variation is ignored, as are fluctuations and correlations. This single effective degree of freedom allows for a description of the system in terms of its fixed point dynamics. The rate of change of the probability $f_+$, say, (which in the limit of large system size is equivalent to the fraction of strong units) with time, is computed by taking into account only the nearest-neighbor synaptic interactions, via the rules defined earlier. The dynamical equation for $f_+(t)$ assumes the following form:

$$\begin{aligned} f_+(t+1) &= r_{+\to+}(t)f_+(t)+r_{-\to+}(t)f_-(t) \\ &\equiv F(f_+(t)) \end{aligned} \tag{1}$$

with the transition probabilities being given by

$$\begin{aligned} r_{+\to+}(t) &= f_+^2(t)+f_-^2(t) \\ &+2f_+(t)f_-(t)(1-p_-(1-p_+)) \\ r_{-\to+}(t) &= 2f_+(t)f_-(t)p_+(1-p_-). \end{aligned} \tag{2}$$

The fractions of strong and weak types are, of course, normalized by definition: $f_+(t)+f_-(t)=1$.

The implicitly time-dependent transition probabilities, which incorporate the effect of nearest-neighbor coupling, introduce non-linearity into the dynamics, an obvious departure from the linear coupled equations of [10]. The deterministic dynamics of Eq. 1 yields stationary states ($f_+^*$) to which the system would relax exponentially starting from an arbitrary initial state. Besides the trivial unstable fixed points at 0 and 1 corresponding to homogeneous, absorbing states (all units being one or the other type), the algebraic equation $f_+^* = F(f_+^*)$ also possesses a stable solution; this is given by

$$f_+^*(p_+,p_-)=\frac{(1-p_-)p_+}{(p_++p_--2p_+p_-)}. \tag{3}$$

(Of course, in the presence of fluctuations, e.g. associated with finite system sizes in mean-field, or in the full solution of the stochastic equations, we would expect the trivial fixed points to be absorbing, and the stable fixed point associated with Eq. 3 to be metastable). The time scale for relaxation to this fixed point is the other dynamically relevant quantity, which again can be extracted from Eq. 1 and is given by

$$\tau=\frac{1}{2}\left[\frac{1}{p_+(1-p_-)}+\frac{1}{p_-(1-p_+)}\right]. \tag{4}$$

This system-level relaxation time is the central quantity with regard to learning and forgetting protocols (see e.g. [10]). It depends on the synapse-level outcome probabilities $p_\pm$, and varies with the location of the corresponding fixed point. It is instructive to illustrate this dependence in the $(p_+,p_-)$ plane, and this has been done in Fig. 2. Such a picture suggests a possible way of defining learning and retention in the coarse-grained representation. To do so, we first define a general time-dependent signal as a 'perturbation' of the system parameters $(p_+,p_-)$ having the following form: $(p_+,p_-)\to(p_++s(t),p_--s(t))$. This choice of signal definition is motivated by taking into account the fact that $p_+-p_-$ plays the role of a 'biasing field'. This has been argued earlier [20] by means of an analogy with spin models; it can also be inferred from the results of applying linear response theory to the original model [35]. Moreover, this way the signal is being applied to both the parameters, rather than preferentially to only one of them. Thus, the application of a signal of this form has the effect of introducing a time dependence into the system parameters.

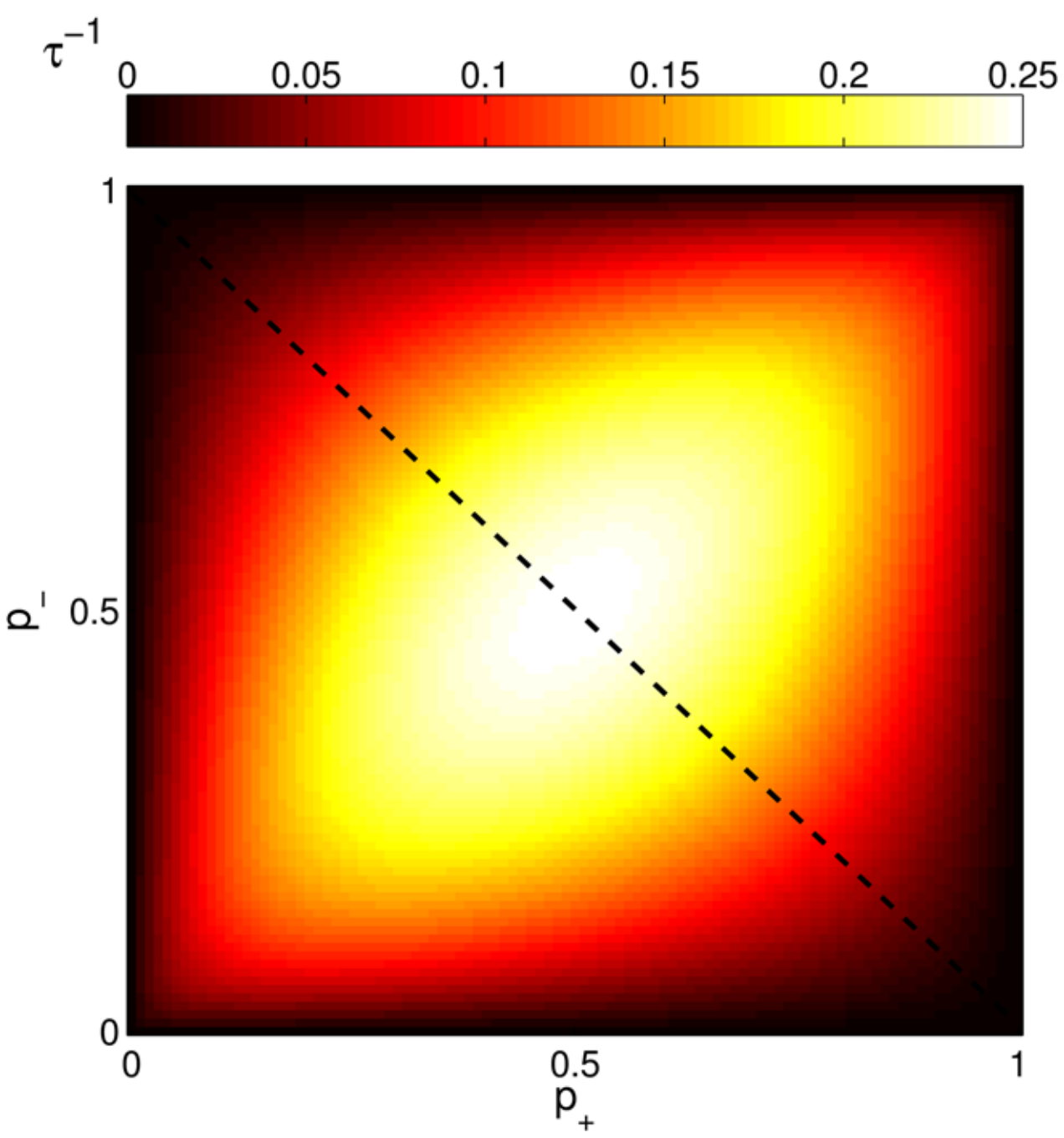

**Figure 2. Relaxation timescales.** Inverse time constant for relaxation ($\tau^{-1}$) as a function of the synaptic parameters $p_\pm$, obtained in the one-dimensional effective representation (see Eq. 4). The dashed line corresponds to the diagonal $p_++p_-=1$, along which the range of imposable signals is maximized. The dark regions near the corners correspond to default configurations with long retention times.
doi:10.1371/journal.pone.0025048.g002

It is easy to work out the consequence of the above signal definition for the simplest example of a constant input signal: the fixed point would shift to a new location along a $p_+ + p_- =$ constant line in the $(p_+, p_-)$ plane. One can, then, imagine a protocol whereby a constant signal is switched on at $t=0$ and persists up to a time $t=T$, following which the system reverts to its original state. Learning and forgetting are both exponential relaxation processes in this setting, and two timescales naturally enter the picture: the learning time constant for moving to the new stable state (after the signal is applied), and the forgetting time constant for reverting to the default fixed point once the signal is turned off. It may be noted that, since the relaxation timescale is a function of the parameters $p_\pm$ of the *end* state, whatever that may be, it is in general different for the learning and the forgetting: the former depends on the values of $p_\pm$ in the presence of the signal, while the latter depends on the unperturbed state.

One of the strengths of the preceding approach is that performance optimization can be directly related to the microscopic parameters $p_\pm$ (in contrast to the approach of [10] where optimization relied on the relative values of multiplicative constants). Fig. 2 suggests an approach to optimizing the performance in the particular scenario under consideration, i.e. achieving long forgetting times and typically shorter learning times: by choosing the default parameters in such a way that the unperturbed state of the system lies near the lower right corner (or the upper left corner), the timescale for retention, being only a function of the unperturbed state, can be made very long, with the average timescale for learning applied signals being shorter. This limit corresponds to having a wide separation between the timescales associated with the two parameters $p_+^{-1}$ and $p_-^{-1}$. (If, alternatively, one were to choose the default values of $p_\pm$ to lie closer to the middle of the graph, the forgetting time constant would be shortened, clearly an undesirable feature.) It may be noted that translating the default state along the diagonal line given by $p_+ + p_- = 1$ only modifies the retention time, while leaving the range of signals that can be absorbed, and thus the *average* learning time, unchanged. Additionally, one observes that given the form of the signal as defined above, which can only produce shifts parallel to the $p_+ + p_- = 1$ diagonal, the range of allowed signals is maximal when the system stays on this diagonal, rather than on any other line parallel to it.

It should be clear from the preceding discussion that we now have a framework in place for studying the responses of the effective system to arbitrary input signals having more general forms of time dependence. We would like to build on the simplest case considered above, and in the next section, carry out a similar exercise of linking system performance to the physically meaningful synaptic parameters for other signal profiles representative of more complex learning protocols. In particular, it would be worthwhile to ascertain if optimal parameter settings can similarly be located in the parameter space of the effective system.

## Analyzing various protocols

We start by reconsidering the first example discussed above, that of a constant signal which is present for a specific duration of time. This signal profile is depicted in Fig. 3(a). The response of the system in terms of the time-dependent relative abundance of strong synapses is illustrated for some different choices of the parameter $p_+ = 1 - p_-$ in Fig. 4; these curves have been obtained by numerically evolving the effective equation for $f_+$. The preceding analysis led us to infer that the system shows long forgetting times (associated with exponential relaxation) when the synaptic time constants are well separated, which is also observable in Fig. 4. Moreover, the effective mean-field representation has the property that the learning time (but not the retention time) shows a dependence on the signal strength as well as the default values of $p_\pm$. Since, in some situations, quick learning is as important as slow forgetting, another quantity of interest is the ratio of forgetting to learning times (which we label as $R$). This quantity will obviously be a function of the signal strength ($s$) too; thus, it is not hard to imagine that for a given system, there will be a particular value of the signal strength that will yield an optimal value for $R$. Continuing to confine ourselves to the $p_+ + p_- = 1$ diagonal (so as to maximize the range of allowable inputs, as mentioned earlier) so that the default system is essentially parametrized by a single variable ($p_+$, say) now, the performance with respect to the parameter $R$ can be visualized, as before, on a two-dimensional $(p_+, s)$ plane. This is shown in the color-coded plots of Figs. 5(a) & (b), which correspond to the analytically computed values and the estimates obtained by numerical simulation (see Methods section for details) respectively; they show hardly any difference. Recall that for any given value of $p_+$, only a certain range of signal values is meaningfully imposable, and this fact is reflected in the phase diagram which does not span the entire range of the $p_+ - s$ plane. While the dependence of $R$ on the value of the signal strength $s$ is readily apparent, it is also clear that close to the extremal values of the parameter $p_+$ (which correspond to having disparate synaptic efficacies) the system does much better *overall*, with higher values of $R$ being attainable in these limits, over a wider range of signal strengths. This result is in agreement with our earlier finding regarding the "functional" benefit of having well-separated synaptic parameters $p_\pm$.

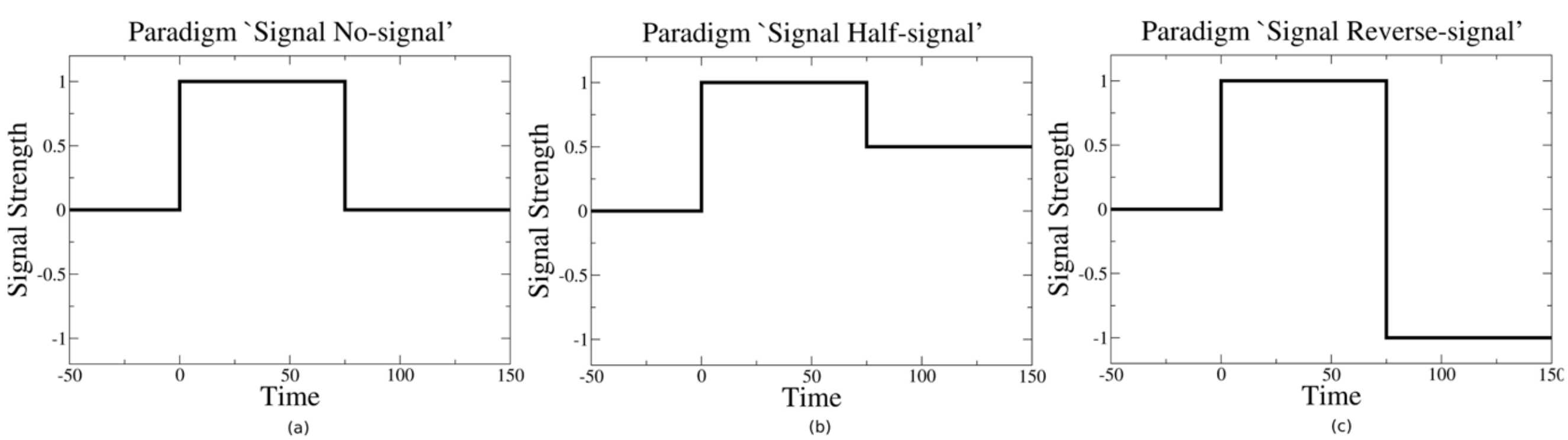

**Figure 3. Temporal forms of the protocols analyzed in the text.** (a) Signal - No Signal (De-adaptation); (b) Signal - Half-Signal (Downscaling); (c) Signal - Reverse Signal (AI).

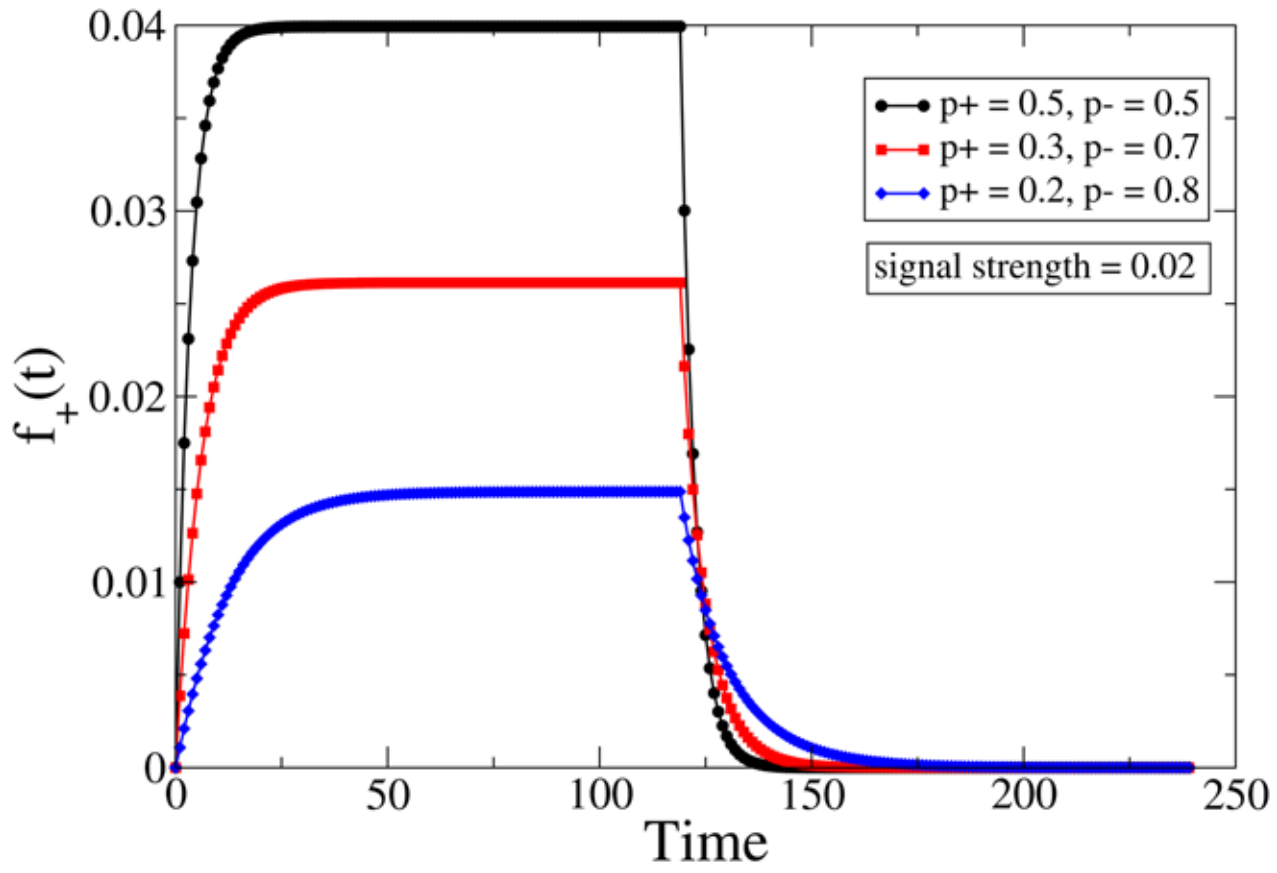

**Figure 4. De-adaptation protocol.** Time dependence of the dynamic variable $f_+$ representing the state of the synaptic population in the effective description, to a signal that is imposed until the system reaches saturation (see Fig. 3(a)). The curves for different settings of $p_\pm$ have been translated vertically to meet the baseline for comparative analysis. The black, red and blue curves correspond to $p_\pm =$ (0.5, 0.5), (0.3, 0.7) and (0.2, 0.8) respectively, and the signal strength is fixed at $s = +0.02$.

The signal example that was just considered also appears in a simulation of deadaptation in the context of the motor adaptation model of [10]. Proceeding along these lines, we next consider two other types of signals, which also are intended to be variants of experimental paradigms considered in [10]. These are referred to as Downscaling and Anterograde Interference (AI), and their forms are shown in Figs. 3(b) & (c).

Let us consider the downscaling signal first. It consists of a phase of constant input, followed by a phase during which the original input is reduced (in our case halved) in magnitude, rather than being completely removed. Just as before, the temporal response of the effective variable $f_+$ to such a signal is illustrated in Fig. 6 with different realizations of the synaptic parameter $p_+$ and some fixed value of signal strength; these curves have been obtained by simulating the effective equation, Eq. 1, by evolving from the default fixed point state. In order to obtain a clearer picture of the variation of the behavior with the basic synaptic parameter as well as the strength of the signal, we consider the corresponding visualization in Figs. 7(a) & (b), which display the analytical and numerical estimates respectively in the allowed region. As before, the two results are extremely similar, although obviously not identical. The quantity that is depicted here is the ratio between the downscaling time (i.e. the timescale associated with relaxing to the fixed point corresponding to the half signal) and the timescale for the initial learning (i.e. the relaxation time to go from the original, unperturbed default state to the fixed point in the presence of the full signal). The joint dependence on the two parameters is brought out in a fairly clear manner, especially to do with the following idea: configurations with $p_\pm$ closer to extremal values have a slower downscaling rate in general, although the performance of any given system certainly depends on the choice of signal strength also.

We next deal with the signal type mimicking the protocol for anterograde interference (refer to Fig. 3(c)). The basic idea here is to apply a constant input signal for a specific duration, and to follow this up by reversing the *sign* of the applied signal, while leaving its magnitude unchanged. Such a scenario was considered in the original context [10] in order to probe the effect that previous learning might have on the subsequent adaptation to an oppositely directed input; in other words, whether past learning could remain imprinted at a deeper level even after appearing to have been erased, thus interfering with the receptiveness to future inputs. For our effective model, the response to such a paradigm is exemplified by the numerically obtained curves in Fig. 8 for three different parameter ($p_+$) choices and a fixed signal strength. Once again, we consider the functional dependence on the synaptic time constants, examining the ratio of the relearning time (for relaxation to the reversed, negative input) to the initial learning time (the timescale for exponential relaxation to the fixed point corresponding to the originally imposed signal), both of which of course also depend on the value of the signal. (For details of the accompanying numerical simulation, see the Methods section.) Figures 9(a) & (b) show a more complex response of the system than we have seen hitherto. It is clear that both the signal strength *and* its orientation need to be factored in; thus for a typical default value of $p_+$ lying in (0,0.5), a positive signal ($s > 0$) shows more anterograde interference than a negative one. (The situation is reversed when we consider $p_+$ in $(0.5,1)$.) This is explained by the observation that a positive signal adds to the strength of the stronger synaptic type, causing more retention of the original signal, and hence a greater time associated with unlearning this to learn an oppositely directed signal.

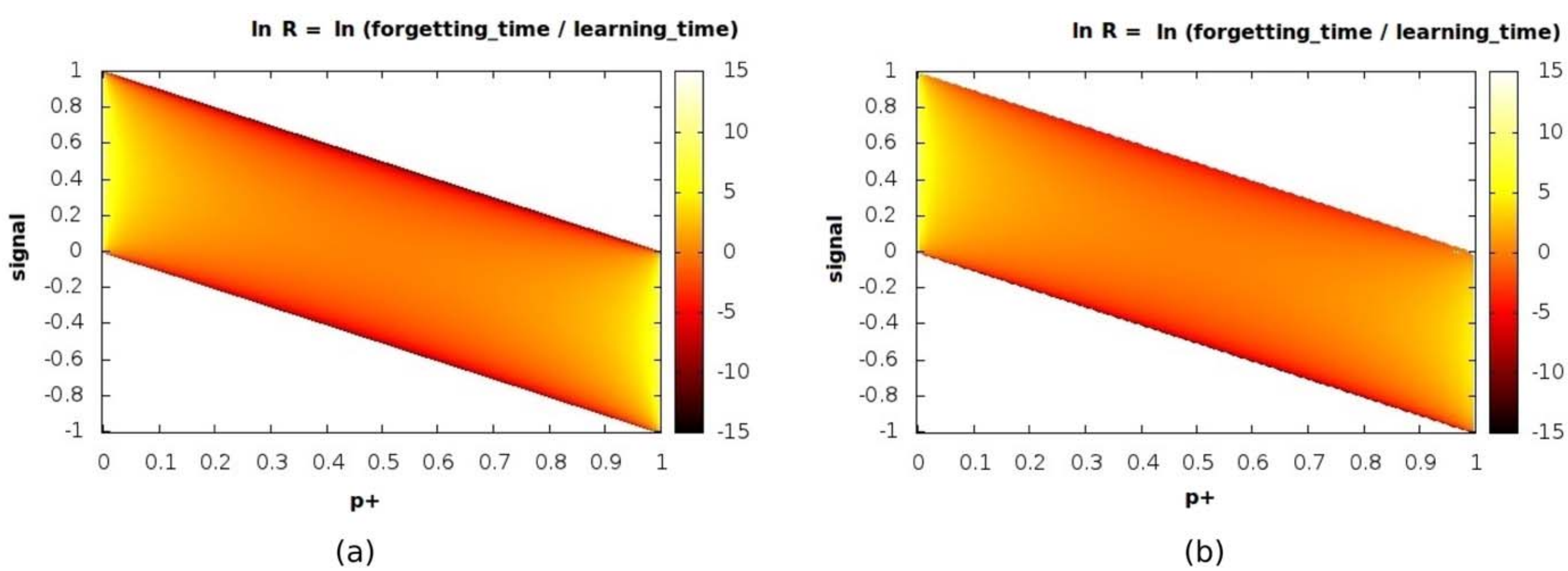

**Figure 5. De-adaptation and synaptic parameters.** Variation of the ratio $R$ of the forgetting timescale to the learning timescale for the Signal - No Signal protocol over the two-dimensional $p_+ - s$ plane ($p_- = 1 - p_+$ has been chosen here). The values of ln $R$ are shown color-coded in the panel to the right of each plot. Both the analytical (a) and numerical (b) estimates are shown.

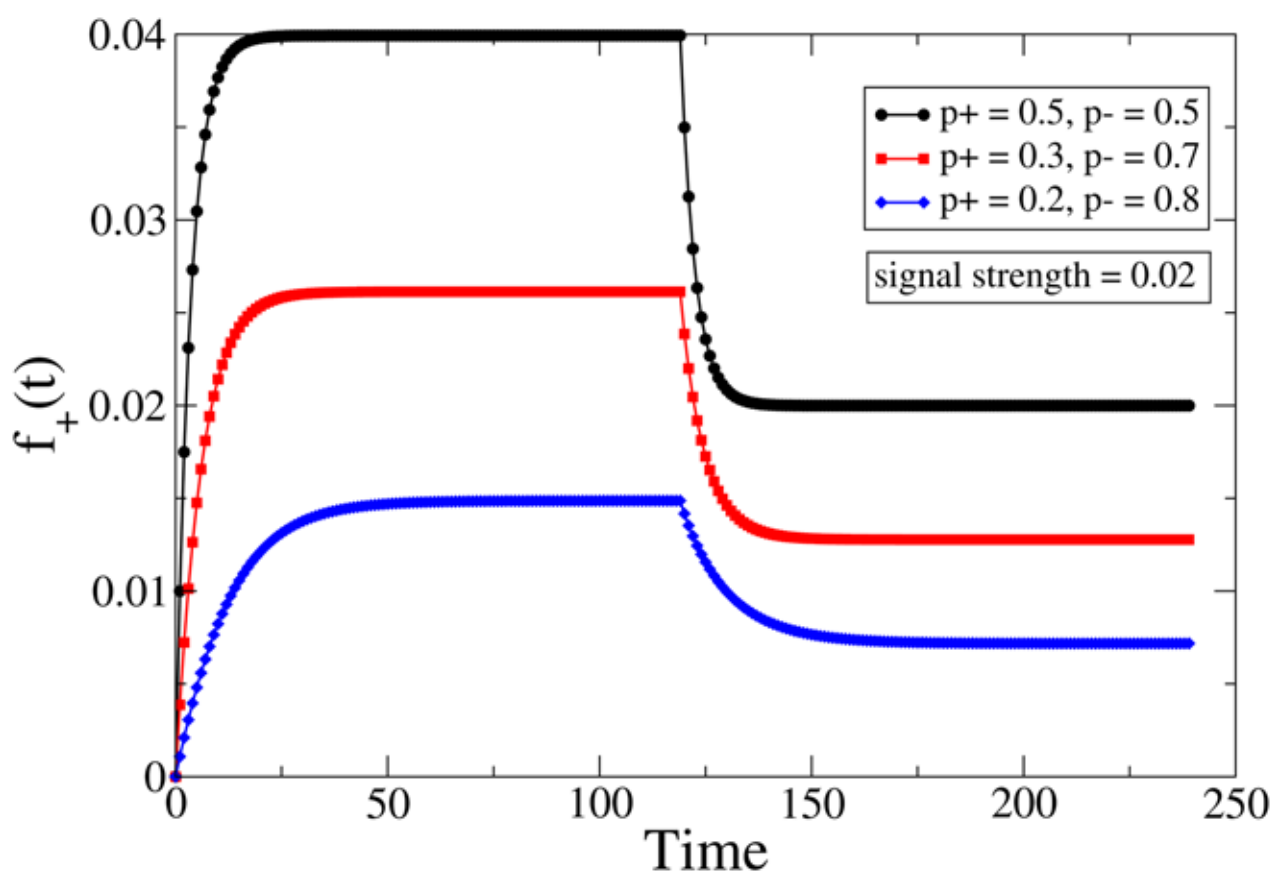

**Figure 6. Downscaling protocol.** Time variation of the effective variable $f_+$ in response to a signal that is reduced by half after a certain period of time (see Fig. 3(b)). The curves for different choices of $p_\pm$ have been translated vertically to meet the baseline for comparative analysis. The black, red and blue curves correspond to $p_\pm =$ (0.5, 0.5), (0.3, 0.7) and (0.2, 0.8) respectively, and the (fixed) signal value is $s = +0.02$.

Our observations on the various protocols have been based on analyses carried out within a mean-field framework, and involve comparisons between timescales of learning and forgetting. These are obtained analytically, as well as by independent simulation, by examining the dynamical relaxation to the fixed points associated with the "bare" system as well as with the system in the presence of a signal. If, therefore, a signal is relearnt before it has been entirely forgotten, it is conceivable that "savings" [10] will be manifested; in particular, this effect may be expected to be quite significant where forgetting is slow, i.e. where $p_+$ and $p_-$ are well separated. This is also consistent with the fact that we get reasonable results for another history-dependent phenomenon, that of anterograde interference, where the system takes longer to learn the reversed signal in the relevant parameter regimes. Our main reason for not probing this further in this paper is that the mean-field approximation, which we have chosen for analytical reasons, is not really the best way to look at the developing correlations associated with history; on the other hand, the full numerical simulation of the exact model will automatically introduce the necessary correlations, and this is a subject that we leave for future work.

## Discussion

A theoretical model such as ours, which introduces game theory-inspired notions of competitive interactions into the field of activity-dependent synaptic plasticity, naturally raises questions about its experimental utility. The main motivation for our model was in fact experimental work concerning motor adaptation [10], where we wished to provide a more microscopic basis to the coupled linear equations presented there. While our equation is formally similar to that presented in [10], the important difference is that it is non-linear, and that microscopic observables, rather than parameters, determine the timescales associated with learning and forgetting. The formal similarity however suggests that many of the protocols can be applied to our mean-field system, with rather similar results to [10], as shown in the last two sections. It suggests also that if correlations are included by solving our full equations to a level that is higher than mean-field, we will be able to incorporate some of the more sophisticated features of the experiments of [10]. The non-linearity of our dynamical equations, and the fact that they show rudiments of network-level memory even in this fairly simple representation, are strong positives. Additionally, learning times in our model scale with the strength of the applied signal, but not the forgetting time, a feature that may actually have some empirical support [36]. This feature introduces an additional dimension of complexity into the dynamics, suggesting for instance, that to every system there corresponds a particular choice of "preferred" signal strength to which it is most receptive, in the sense of learning it the quickest (for an extension of this to the phenomenon of hearing, see [37]). More importantly, the present analytical approach provides a rather transparent link between the quantities characterizing the system performance in this case the various relaxation rates and those defined at the fundamental micro-level of neurons and synapses ($p_\pm$ and $s$). This obviates the need to have to fish around for optimal parameters, and provides a better handle on matters having to do with the locating of the right synaptic parameters corresponding to a particular behavior.

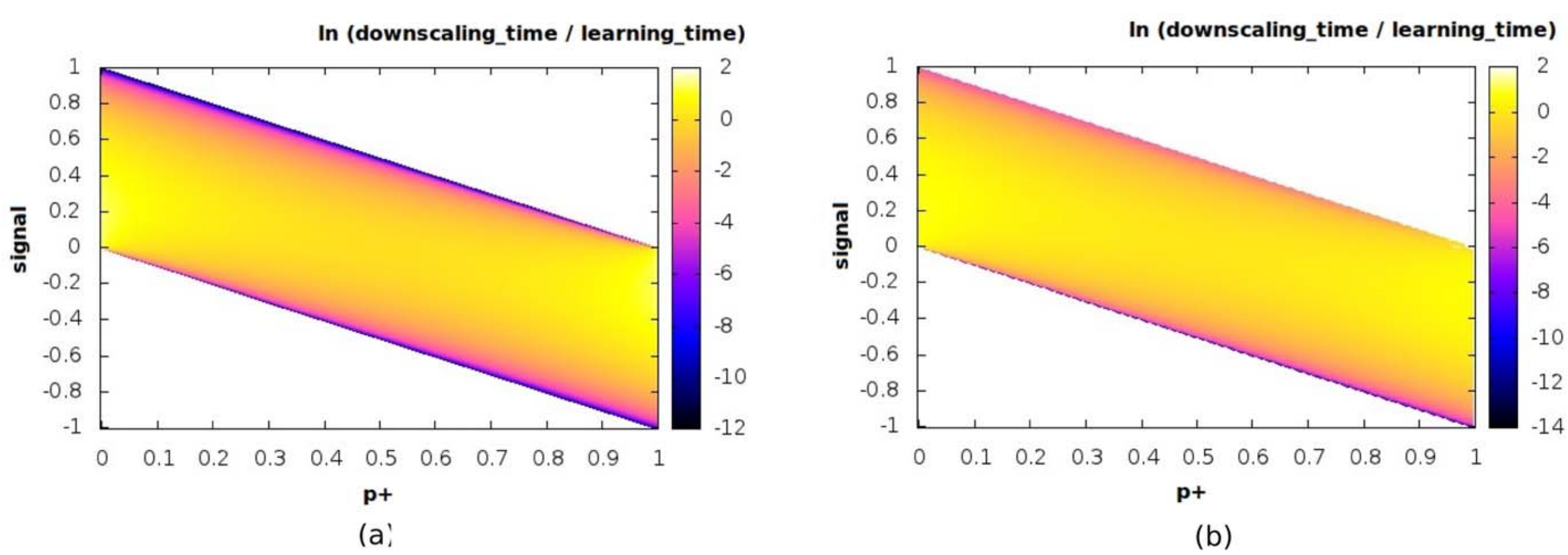

**Figure 7. Downscaling and synaptic parameters.** Variation of the ratio of the downscaling timescale to the timescale of initial learning for the Signal - Half Signal protocol over the two-dimensional $p_+ - s$ plane ($p_-$ is set equal to $1-p_+$). The values of the logarithm of the ratio are shown color-coded in the panel to the right of each plot. Both the analytical (a) and numerical (b) estimates are shown.

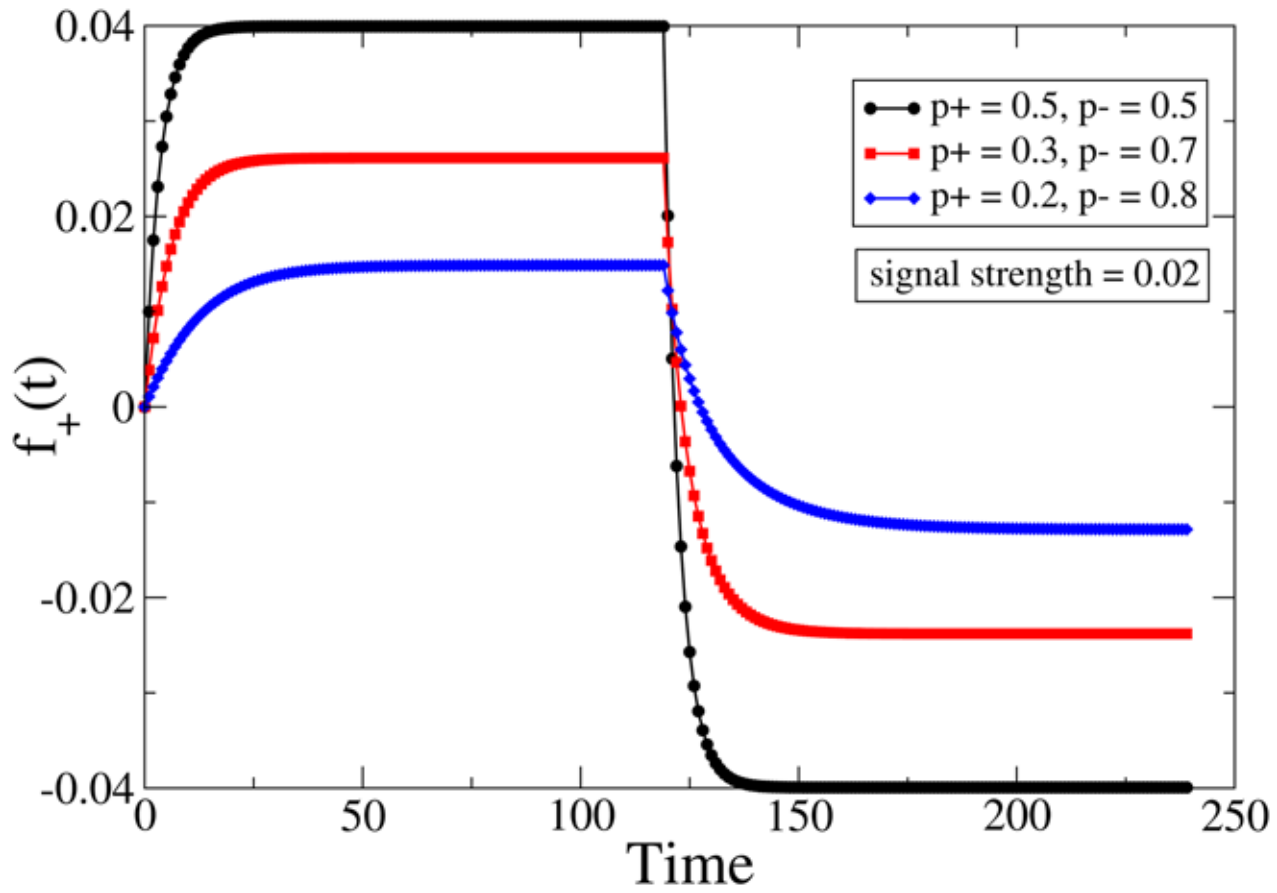

**Figure 8. Anterograde interference.** Response of the synaptic population variable $f_+$ to the application of a signal followed by its reversal (see Fig. 3(c)). The curves for different choices of $p_\pm$ have been translated vertically to meet the baseline for comparative analysis. The black, red and blue curves correspond to $p_\pm =$ (0.5, 0.5), (0.3, 0.7) and (0.2, 0.8) respectively, and the (fixed) signal value is $s = +0.02$.

Referring back to the plots of the previous section for the protocols considered, specifically those corresponding to de-adaptation and downscaling, we note that the choice of disparate synaptic timescales (directly relatable to the synaptic weights) is in general associated with more favorable performance, in the sense of efficient learning and longer retention, even when signal strengths are variable. This broad conclusion is consistent with the findings of ref. [10], whose model consisting of coupled fast and slow components also shows markedly improved agreement with the results of hand-reaching experiments, in comparison with its single-timescale competitors.

Before concluding, we mention some possible extensions of this work. One can, of course, go beyond mean field, retaining spatial information, and correlations, in the determination of the neuronal outcomes. The effect of doing so would be to introduce additional degrees of freedom, and corresponding timescales, into the description of the system dynamics. In a pair-approximation [20] for instance, which takes into account the correlations between nearest neighbors, the macro-level dynamics would contain *two* relaxation rates, and show richer behavior. Thus, working with such a higher-order effective representation might lead to a more fruitful link with experiment. An interesting possibility would also be to look at directed synapses, which would in fact amount to changing the update rules [35] of the present model. Also, we would like to incorporate ideas from spike-timing-dependent plasticity [38] into our model, to give some of our timescales a firmer microscopic underpinning. Last but by no means least, the inclusion of our model synapses into realistic networks is a major goal, to put our work in the context of recent models of brain learning [39]; the use of game theory in evolving networks [40] would be of particular use in this endeavour.

To summarize, we have explored a novel model for synaptic plasticity, sketched in earlier work [23], which incorporates a notion of competition (between synapses in their distinct states) appropriate to decision-making paradigms in fluctuating circumstances. An approximate, coarse-grained description for obtaining learning behavior has been explored at length. While highlighting interesting features, including the effect of choosing dissimilar timescales along the lines of a body of previous work, our approach also suggests that more involved quantitative treatment may lead to more concrete connections with experiment.

## Methods

The effective equation Eq. 1 is evolved numerically in the presence of the chosen signal type $s(t)$. For every simulation, the starting value of $f_+$ is assumed to correspond to the fixed point of the default, unperturbed system. Once a signal $s$ is turned on, the system is allowed to evolve from its initial state until the time that it "saturates" at a new state; this is implemented by imposing the condition that the fractional change in $f_+$ over one time-step be less than a certain tolerance limit (chosen to be equal to $10^{-8}$ here). The signal is assumed to remain applied until this condition is reached, at which point the system may be considered to have "learnt" the signal. Thereafter the signal is removed or changed to a new value depending on the protocol. The time taken to reach saturation, beginning from the moment the signal was switched on, is taken to be the timescale for the corresponding process.

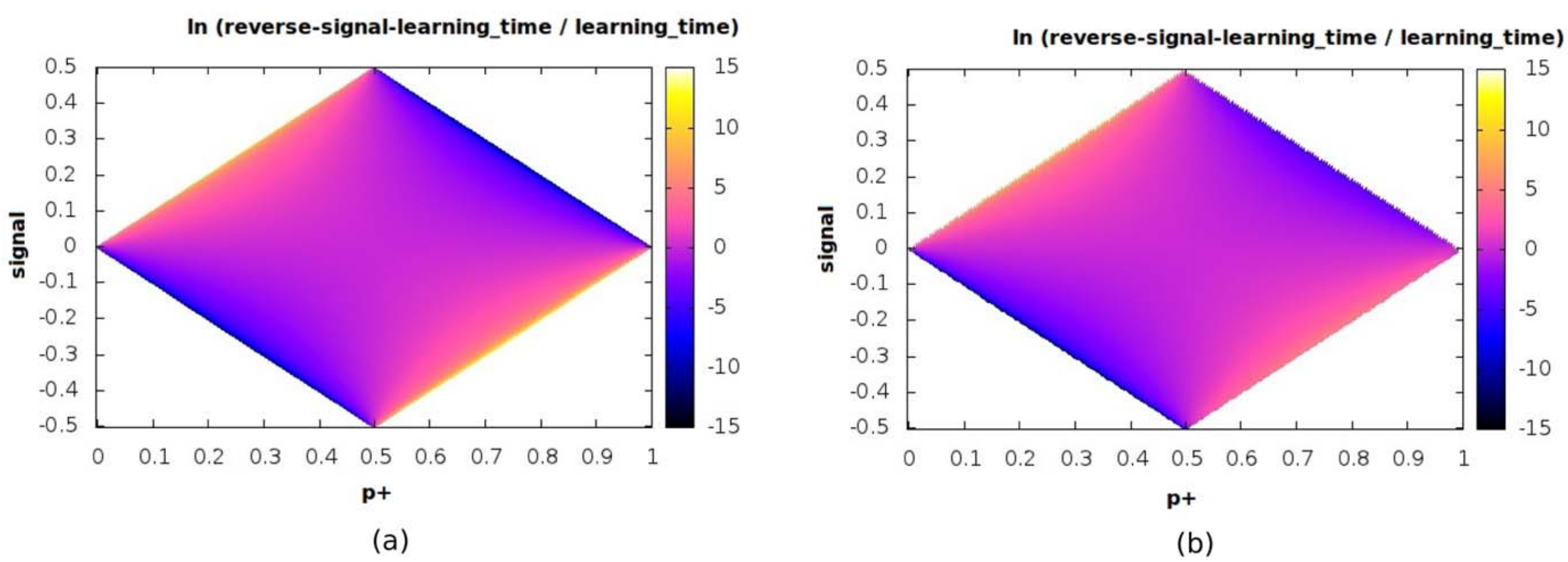

**Figure 9. Anterograde interference and synaptic parameters.** Variation of the ratio of the timescale for learning the reversed signal to the timescale for initial learning for the Signal - Reversed Signal protocol over the two-dimensional $p_+ - s$ plane ($p_-$ is set equal to $1-p_+$). The values of the logarithm of the ratio are shown color-coded in the panel to the right of each plot. Both analytical (a) and numerical (b) estimates are displayed.

In the de-adaptation and downscaling simulations, the determination of both the involved timescales (initial learning followed by forgetting or downscaling respectively) is straightforward and follows the procedure outlined above. The third protocol (signal - reversed signal) involves reversing the sign of an applied signal, and estimating the time for learning this reversed signal. For obtaining this time, the system is first allowed to evolve back from the stable state that was attained in the presence of the initially applied signal, to the *default level* (i.e. the unperturbed system state). The time for learning the reversed signal is only measured from this point onwards, until $f_+$ saturates at the fixed point corresponding to the reversed signal (following the procedure of the previous paragraph). In other words, the system is first allowed to "forget" the initially imposed signal (by getting back to its default state), and then made to "learn" the sign-reversed counterpart.

## Acknowledgments

AM thanks the Institut de Physique Théorique, where part of this work was carried out, for its hospitality.

## Author Contributions

Conceived and designed the experiments: GM AM. Performed the experiments: AAB GM. Wrote the paper: AAB GM AM.